\documentclass{article}

\let\shortcite\cite
\usepackage{algorithm}
\usepackage{algorithmic}
\usepackage{pifont}
\usepackage{latexsym}
\usepackage{amsfonts}
\usepackage{amsmath}
\usepackage{amssymb}
\usepackage{yhmath}
\usepackage{graphicx}
\newcommand{\dist}[1]{\ensuremath{{\it d}(#1)}}

\usepackage{natbib}
\let\cite\citep

\usepackage{pifont}
\newcommand{\p}{{\rm P}}
\newcommand{\np}{{\rm NP}}
\newcommand{\conp}{{\rm coNP}}

\newcommand{\sigmatwo}{\ensuremath{\Sigma_2^{p}}}
\newcommand{\pitwo}{\ensuremath{\Pi_2^{p}}}

\let\shortcite\cite
\newtheorem{theorem}{Theorem}
\newtheorem{corollary}[theorem]{Corollary}

\newtheorem{lemma}[theorem]{Lemma}

\newcommand\qedblob{\ding{113}}
\def\literalqed{{\ \nolinebreak\hfill\mbox{\qedblob\quad}}}

\newtheorem{observation}[theorem]{Observation}
\newtheorem{example}[theorem]{Example}

\newenvironment{proofs}{{\it Proof.}\hspace*{1em}}{\literalqed\smallskip} %

\newenvironment{proofsketch}{{\it Proof Sketch.}\hspace*{1em}}
{\literalqed\smallskip} %

\newcommand{\prob}[3]{
\begin{description}
  \item[Name:] #1
  \item[Given:] #2
  \item[Question:] #3
\end{description}}
\newcommand{\score}[1]{\ensuremath{{\rm score}(#1)}}

\title{Kemeny Consensus Complexity}

\author{
Zack Fitzsimmons\\
 Dept.\ of Math.\ and Computer Science\\
 College of the Holy Cross\\
Worcester, MA 01610 \and
  Edith Hemaspaandra\\
  Department of Computer Science\\
  Rochester Institute of Technology \\
  Rochester, NY 14623}%

\date{May 18, 2021} %

\begin{document}
\sloppy

\maketitle

\begin{abstract}

The computational study of election problems generally focuses on questions related to the winner or set of winners of an election. But social preference functions such as Kemeny rule output a full ranking of the candidates (a consensus). We study the complexity of consensus-related questions, with a particular focus on Kemeny and its qualitative version Slater. The simplest of these questions is the problem of determining whether a ranking is a consensus, and we show that this problem is coNP-complete. We also study the natural question of the complexity of manipulative actions that have a specific consensus as a goal. Though determining whether a ranking is a Kemeny consensus is hard, the optimal action for manipulators is to simply vote their desired consensus. We provide evidence that this simplicity is caused by the combination of election system (Kemeny), manipulative action (manipulation), and manipulative goal (consensus). In the process we provide the first completeness results at the second level of the polynomial hierarchy for electoral manipulation and for optimal solution recognition.
\end{abstract}

\section{Introduction}

Elections are a widely used tool for aggregating the preferences of
several agents into a collective decision. Often the goal is to determine
a single winner or set of winners from among a set of candidates.
However, in other cases, such as constructing a meta-search engine~\cite{dwo-kum-nao-siv:c:rank-aggregation} or genetic maps~\cite{jac-sch-alu:j:kemeny-bioinformatics},
the natural desired outcome is a ranking. %

One of the most compelling ways of aggregating preferences is the Kemeny rule. It is known that computing a Kemeny consensus (i.e., a ranking closest to the electorate) is a computationally difficult problem.
We show that even simply checking if a given ranking is a consensus is coNP-complete. This
problem is naturally motivated by an
agent wanting to verify the claimed outcome of an election.

One of the most important lines of research in the
computational study of elections (see, e.g.,~\citet{fal-rot:b:handbook-comsoc-control-and-bribery}) is the study
of different manipulative actions
such as manipulation and control~\cite{bar-tov-tri:j:manipulating,bar-tov-tri:j:control}, where
an agent (or agents) seek to ensure their preferred outcome by either
voting strategically or modifying the structure of the election.
In each of these models, the goal is typically to ensure a
preferred candidate wins. For scenarios where the collective decision
 is a consensus, it is natural to
consider manipulative actions where the goal of the
agent(s) is to reach a preferred consensus.

Even though the problem of determining whether a ranking is a Kemeny consensus is hard, the optimal action for the manipulators is to simply vote their desired consensus. We provide evidence that this simplicity is caused by the combination of the manipulative action (manipulation), the manipulative goal (consensus), and the election system (Kemeny).
In particular:
\begin{itemize}
    \item Determining if a given ranking is a Kemeny consensus is coNP-complete. (Section~\ref{s:KCR})
    \item Control by deleting candidates for Kemeny with the goal of a particular consensus is \sigmatwo-complete (and thus, unlike manipulation, the optimal control action is not polynomial-time computable). (Section~\ref{sec:control})
    \item We provide evidence that manipulation (to winner) for Kemeny is also much harder than manipulation to consensus,
    by showing that manipulation (to winner) for a natural variant of Slater (the qualitative version of Kemeny) is \sigmatwo-complete. (Section~\ref{sec:manipulation})
    
    \item The choice of system matters as well. For example the
    optimal action for the manipulators to reach a consensus is not polynomial-time computable for Borda. (Section~\ref{sec:borda})
\end{itemize}

\section{Preliminaries}

An election consists of a set of candidates $C$ and
a collection of voters $V$ where each voter has a ranking (total order
preference) over the set of candidates. For example, $a > b > c$, where $>$ denotes strict preference,  is a vote over
$\{a,b,c\}$.

We consider voting rules that are social preference functions, which map an election to a set of one or more rankings (total orderings) of the candidates. One of the most-important
social preference functions is the Kemeny rule~\cite{kem:j:no-numbers}.

A ranking $>$ is a Kemeny consensus if the sum of the Kendall tau distances to the voters is minimal,
i.e., $\sum_{a > b} N(b,a)$ is minimal, where for candidates $a$ and
$b$, $N(b,a)$ denotes the number of voters that state $b > a$.

It is often useful to refer to the induced weighted majority graph of the election when working with the Kemeny rule. The weighted majority
graph of an election $(C,V)$ has a vertex for each candidate and for each pair of candidates $a,b \in C$ if $N(a,b) > N(b,a)$ there is an arc $(a,b)$ labeled with $N(a,b) - N(b,a)$.

\begin{example}\label{ex:kemeny}
Consider an election with candidates $\{a,b,c,d\}$ and
three voters with their votes, and the corresponding induced
weighted majority graph below.
\begin{minipage}[c]{0.25\textwidth}%
    \hspace{\textwidth}
\end{minipage}%
\begin{minipage}[c]{0.25\textwidth}%
 \begin{itemize}
    \item $a > b > c > d$
    \item $c > a > d > b$
    \item $b > c > d > a$
\end{itemize}
\end{minipage}%
\begin{minipage}[c]{0.5\textwidth}%
\includegraphics[scale=0.75]{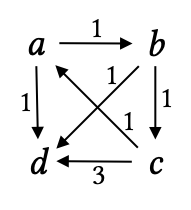}
\end{minipage}

Consensuses: \ $a > b > c > d$, \ \ $b > c > a > d$, \ \ $c > a > b > d$. Kendall tau distance: 6. 
\end{example}

We consider several different computational problems relating to
the Kemeny rule. For readability, the formal definitions of these problems
are deferred to where the results appear.

We assume that the reader is familiar with the complexity classes \p, \np,
and \conp.
Our complexity results also concern the class \sigmatwo = $\np^\np$, a
class at the second level of the polynomial hierarchy, which is the
class of problems
solvable by an \np-machine with access to an \np\ oracle~\cite{mey-sto:c:reg-exp-needs-exp-space,sto:j:poly}.

\section{Consensus Recognition}
\label{s:KCR}
We now formally define the problem of determining whether a ranking is a Kemeny consensus.

\prob{Kemeny Consensus Recognition}%
{An election $(C,V)$ and a total order $X$.}%
{Is $X$ a Kemeny consensus of the election?}

\citet{hud:j:kemeny-complexity} observes that
the Kemeny Consensus Recognition problem (there called Order Recognition)
is in coNP, and
conjectures it is coNP-complete.\footnote{Hudry uses Turing reductions. We look at the standard notion of polynomial-time many-one reductions, which gives stronger results.}
Since it is easier to think about NP than about coNP, we will often look at the complement, i.e., determining whether $X$ is not a consensus.
As usual, the upper bound is easy to see: Note that $X$ is not a Kemeny consensus if and only if there exists a total order whose distance to the election is less than that of $X$.

Also note that the Kemeny Consensus Recognition problem is not in NP unless NP = coNP, since the Kemeny score of an election (Kendall tau distance to a consensus) is greater than $k$ if and only if there exists a total order that is a Kemeny consensus whose score is greater than $k$. So if Kemeny Consensus Recognition is in NP, then determining if the Kemeny score of an election is greater than $k$ is in NP. This latter problem is coNP-complete, since it is in essence the complement of the problem Kemeny Score, which is NP-complete~\cite{bar-tov-tri:j:who-won}.

The above does not imply that Kemeny Consensus Recognition is coNP-hard. It merely says that, assuming NP $\neq$ coNP, the problem is in coNP $-$ NP. Under the assumption that NP $\neq$ coNP, there are problems in coNP $-$ NP that are not coNP-complete~\cite{lad:j:np-incomplete}.
A natural candidate of such a problem is graph nonisomorphism problem. Note that this problem has some ``easiness'' properties that are not shared by any natural coNP-complete problem, such as a zero-knowledge proof for the complement~\cite{gol-mic-wig:j:zero-knowledge} and a quasi-polynomial time algorithm~\cite{Babai}.

We will now prove Hudry's conjecture that Kemeny Consensus Recognition is coNP-complete (as Theorem~\ref{t:KCR}).

Optimal solution recognition problems induced by optimization problems are very natural decision problems, but there are only a couple of results in the literature. \citet[Theorem 5]{pap-ste:j:tsp} show that Minimum TSP Tour Recognition is coNP-complete. \citet{arm-jac:j:global-verification} study the global verification problem (which is the complement of optimal solution recognition) related to various NP optimization problems and show that the optimal solution recognition problems for Vertex Cover, MAX-SAT, and MAX-$k$-SAT ($k \geq 2$) are each coNP-complete.

Our proof of Theorem~\ref{t:KCR} will use the coNP-completeness of Minimum Vertex Cover Recognition. %

\prob{Minimum Vertex Cover Recognition}%
{A graph $G$ and a set of vertices $X$.}%
{Is $X$ a minimum vertex cover of $G$?}

\begin{theorem}[\cite{arm-jac:j:global-verification}]
Minimum Vertex Cover Recognition is \conp-complete.
\end{theorem}

The Kemeny Score problem was shown hard by a reduction from Feedback Arc Set (FAS)~\cite{bar-tov-tri:j:who-won} and the proof shows that these problems are very closely related (made precise
in the statement of Lemma~\ref{l:faskc}).

We will next show that the following problem is coNP-hard. 

\prob{Minimum FAS Recognition}%
{An irreflexive and antisymmetric directed graph $G$ and a set of arcs $X$.}%
{Is $X$ a minimum fas of $G$ (a minimum set of arcs such that
$G - X$ is acyclic)?}

\begin{theorem}
\label{t:minfas}
Minimum FAS Recognition is \conp-complete.
\end{theorem}

\def\minfasproof{%
\begin{proofs}
We will reduce from Minimum Vertex Cover Recognition. Let $G$ be a graph and $X$ a set of vertices of $G$. Now define directed graph
$\widehat{G}$ from $G$ as in the construction of the~\citet{kar:b:reducibilities} reduction
from Vertex Cover to FAS, i.e., 
\begin{itemize} 
\item
$V(\widehat{G}) = \{v,v' \ | \ v \in V(G)\}$, and
\item $A(\widehat{G}) = \{(v,v') \ | \ v \in V(G)\} \cup \{(v',w), (w',v) \ |$ $\  \{v,w\} \in E(G)\}$.
\end{itemize}

Let $\widehat{X} = \{(v,v') \ | \ v \in X\}$. It follows from the proof of the reduction from~\citet{kar:b:reducibilities} that $X$ is a vertex cover of $G$ if and only if $\widehat{X}$ is a fas of $\widehat{G}$ and that $X$ is of minimal size if and only if $\widehat{X}$ is of minimal size. This completes the reduction.~\end{proofs}}

\minfasproof

As mentioned above, feedback arc sets and Kemeny consensuses are very closely related~\cite{bar-tov-tri:j:who-won}.
We need a slightly unusual formulation of this relationship.

\begin{lemma}
\label{l:faskc}
For $G$ a directed graph, let $e(G)$ be the election with candidates $V(G)$ and for each arc $(a,b) \in A(G)$ one voter voting $a > b$ followed by all candidates in $V(G) - \{a,b\}$ in lexicographical order and one voter voting all candidates in $V(G) - \{a,b\}$ in reverse lexicographical order followed by $a > b$. This election is computable in polynomial time, and has $G$ with all arc weights 2 as its induced weighted  majority graph~\cite{mcg:j:election-graph}.

For $X$ a {\em minimal} fas of $G$ (i.e., $X$ is a fas of $G$ and no strict subset of $X$ is a fas), and $\widehat{X}$ a total order consistent with $G - X$ (i.e., if $(a,b) \in A(G) - X$, then $a > b$ in $\widehat{X}$), it holds that $X$ is a {\em minimum} fas if and only if $\widehat{X}$ is a Kemeny consensus of $e(G)$.
\end{lemma}

This gives us a reduction from Minimum FAS Recognition to Kemeny Consensus Recognition, which
gives us the following theorem.

\begin{theorem}
\label{t:KCR}
Kemeny Consensus Recognition is \conp-complete.
\end{theorem}

\begin{proofs}
Given $G$ and $X$, if $X$ is not a minimal fas (which can be determined in polynomial time),
then output something that is not an instance of the problem. If $X$ is a minimal fas, then output $e(G)$ (as defined in Lemma~\ref{l:faskc}) and a total order consistent with $G - X$ (which can be computed in polynomial time, since $G - X$ is
acyclic).~\end{proofs}

From the above, one might think that coNP-hardness for an optimal solution recognition problem follows from a straightforward modification of the reduction for the related NP-complete decision problem. But this is only the case when the witnesses of the two decision problems directly correspond to each other. This is usually not the case. See for example the proof of the analogous result for tournaments later in this paper~(Theorem~\ref{t:CRT}).

\section{Manipulative-Actions-to-Consensus}
\label{s:actions}

In the previous section we showed that Kemeny Consensus Recognition is coNP-complete. Given the hardness of this problem, does it follow that manipulative actions with the goal to reach a specific consensus are hard?
This is true if we look at decision problems such as ``Given an election and a total order $X$, can we perform a manipulative action such that $X$ is a consensus.'' Such decision problems typically inherit the coNP-hardness (e.g., by having no manipulators). It is still interesting to look at these decision problems, since they may be complete for classes above coNP, which limits the tools we have to solve these problems. Standard approaches for solving problems in \np\ or \conp\ such
as using SAT solvers are not appropriate for solving problems that are complete for higher levels of the polynomial hierarchy such as \sigmatwo{}.

We will also look at the problem of determining the manipulative action. It is possible that it is easy to determine the best action, even though it is hard to determine whether such an action leads to the desired outcome. In fact:

\begin{observation}
\label{o:km}
{
Consider Kemeny-Manipulation-to-Consensus, in which we are given an election, a collection of manipulators, and a desired consensus $X$, and we ask if the manipulators can vote so that $X$ is a Kemeny consensus of the resulting election.
It is easy to see that 
a total order $X$ can be made a consensus if and only if $X$ is a consensus when all manipulators vote $X$ (for details see the appendix).

And so the optimal action for the manipulators is straightforward, namely to vote $X$, and
the complexity of the associated decision problem Kemeny-Manipulation-to-Consensus is the same as for the recognition problem, namely, \conp-complete.}
\end{observation}

Now we ask: What makes it easy to determine the manipulative action?  Is it the election system (Kemeny)? Is it the manipulative action (manipulation)? Is it the manipulative goal (consensus)?

Note that the observation above has interesting repercussions for other manipulative actions and for other manipulative goals. For example, in bribery, we can assume that all bribed voters vote the same $X$, where $X$ is a consensus after bribery. And if the goal of the manipulators is to make a preferred candidate $p$ a winner, we can assume that all manipulators vote the same $X$, where $X$ is a consensus after manipulation. (Since if there is a manipulation such that $p$ is a winner, then there is a manipulation with a consensus $X$ that ranks $p$ first. But then $X$ is also a consensus when all manipulators vote $X$.)

Despite this simplicity of all manipulators/bribed voters voting the same, we will provide evidence in the next couple of sections that determining the optimal manipulation to obtain a Kemeny consensus
is easy because of the combination of election system (Kemeny), manipulative action (manipulation), and manipulative goal (consensus).

\section{Control-to-Consensus}
\label{sec:control}
Electoral control models whether the structure of an election can be modified to ensure a preferred outcome~\cite{bar-tov-tri:j:control}.
Control(-to-Winner) problems for Kemeny tend to be $\sigmatwo$-complete~\cite{fit-hem-hoo-nar:c:vhc}
(note that winner determination is already complete for parallel access to \np~\cite{hem-spa-vog:j:kemeny}). In this section we provide evidence
that this is also the case for Control-to-Consensus. Note that this implies that, unless NP = coNP, the optimal control action to obtain a Kemeny consensus is not polynomial-time computable (in contrast to manipulation).

$\sigmatwo$ lower bounds are often hard to prove, in part because there are fewer known $\sigmatwo$-complete problems (see~\citet{sch-uma:j:PH-part-one} for a list)
and also because one needs a closer correspondence between the two problems than for NP-hardness reductions.

We first show that optimal solution recognition for the \sigmatwo-complete problem Generalized Node Deletion (GND)~\citep{rut:j:prop-truth-maintenance} is
$\pitwo$-complete.

\prob{Minimum GND Recognition}
{A graph $G$, integer $\ell$, and set of vertices $X$.}%
{Is $X$ a minimum set of vertices such that $G - X$ does not contain $K_{\ell+1}$ (a
clique of size $\ell+1$)?}

\begin{theorem}\label{t:mgndr}
Minimum GND Recognition is $\pitwo$-complete.
\end{theorem}

This is the first completeness result at the second level of the polynomial hierarchy for optimal solution recognition.
For details, see the appendix.

The natural deletion analogues of Minimum Vertex Cover (resp.\ FAS) Recognition
where we are additionally given a delete limit $k$ and ask if there exists a set
of at most $W$ vertices such that $X$ is a minimum vertex cover (minimum fas, respectively) of $G - W$
are also $\sigmatwo$-complete (see appendix). Since there is a dearth of natural $\sigmatwo$-complete problems, these results are interesting in their own right. 

We will now look at control by deleting candidates (CDC).  We will show that the following problem is $\sigmatwo$-complete.

\prob{Kemeny-CDC-to-Consensus}%
{An election $(C,V)$, delete limit $k$, and a total order $X$ over $C$.}%
{Does there exist a set $D \subseteq C$ of at most $k$ candidates such that $X$ restricted to $C - D$ is a Kemeny consensus of $(C-D,V)$?}

Though this problem is not the most natural, it does provide evidence that Kemeny-Control-to-Consensus problems are $\sigmatwo$-complete.

\begin{theorem}\label{t:kcdcc}
Kemeny-CDC-to-Consensus is $\sigmatwo$-complete.
\end{theorem}

Note that in the definition of Kemeny-CDC-to-Consensus, it is important that $X$ is a total order over $C$. If it were over $C-D$, we would be able to see which candidates are deleted from the problem instance (and the problem would be equivalent to Kemeny Consensus Recognition). However, this makes the problem different from the $\sigmatwo$-complete FAS problem, since total order $X$ ranks all candidates. 
This means that the straightforward $\sigmatwo$ analogue of the reduction from Minimum FAS Recognition to Kemeny Consensus Recognition from the proof of 
Theorem~\ref{t:KCR} does not work. In that reduction, the order was a total order consistent with the directed acyclic graph $G - X$, where $X$ is a fas. However, before deletion, $G-X$ is not necessarily acyclic! 
To prove $\sigmatwo$-completeness of Kemeny-CDC-to-Consensus, we need different, less natural $\sigmatwo$-complete versions of Vertex Cover and Feedback Arc Set Recognition that look more like Kemeny-CDC-to-Consensus. In particular, we need to make sure that the solutions for Vertex Cover and Feedback Arc Set are (not necessarily optimal) solutions for the whole graph. Details can be found in the appendix.

There are other types of control, most notably control by adding candidates and control by adding/deleting voters. As problems, these are more compelling. For example, the definition of control by deleting voters (CDV) to consensus is straightforward and natural.

\prob{Kemeny-CDV-to-Consensus}%
{An election $(C,V)$, delete limit $k$, and a total order~$X$.}%
{Does there exist a set $W \subseteq V$ of at most $k$ voters such that $X$ is a Kemeny consensus of $(C,V-W)$?} 

One might think that, in analogy to optimal action for manipulators being voting the consensus, the optimal action for CDV would be to simply delete voters furthest from the desired consensus (and for CAV to simply add voters closest to the desired consensus). However, the following example shows that this is not the case.

\begin{example}
Consider an election with candidates $\{a,b,c\}$,
five voters: three voting $a > b > c$, one voting $a > c > b$,
and one voting $c > b > a$,
delete limit 1, and desired consensus $a > c > b$.

Note that $a > c > b$ is not a consensus. 
If we delete the voter furthest from the consensus (i.e., the voter voting $c > b > a$) then $a > c > b$ is not a consensus, but if we delete one of the $a > b > c$ voters then $a > c  > b$ is a consensus.
\end{example}

This example with one of the $a > b > c$ voters and the $c > b > a$ voter
as the unregistered voters %
and an add limit of 1 shows
the analogous counterexample for Kemeny-CAV-to-Consensus.

We conjecture that all these control-to-consensus problems are $\sigmatwo$-complete. 
However, we cannot modify the approach above in a simple way, since one arc in a graph does not correspond to one voter in the corresponding election. This is also the reason that the complexity of ``regular'' Kemeny voter control(-to-winner)
is still open~\cite{fit-hem-hoo-nar:c:vhc}.

\section{Manipulation(-to-Winner)}
\label{sec:manipulation}

Showing that manipulation is hard is hard! For example, it is not too hard to show that control for Borda is hard~\cite{rus:t:borda}, but the complexity of (coalitional) manipulation for Borda
was open for a long time and NP-completeness was shown only after discovering an appropriate NP-complete problem in the scheduling literature~\cite{dav-kat-nar-wal-xia:j:borda-manip,bet-nei-woe:c:board-manip}. And proving the NP-completeness of manipulation for Copeland$^\alpha$ for $\alpha \neq 0.5$ involved construction of elaborate gadgets~\cite{fal-hem-sch:c:copeland-ties-matter,fal-hem-sch:c:copeland01}.

The reason that it is so hard to prove manipulation hard is that the manipulators do not follow any structure other than voting a total order. This means that basically all the structure needs to come from the nonmanipulators.

For Kemeny, we know from Section~\ref{s:actions} that we can assume that all manipulators vote the same. So all we have to work with is one total order. Though we conjecture that Kemeny-Manipulation is $\sigmatwo$-complete, we have not succeeded in proving this. The closest we got is the following theorem, which is explained in more detail after the theorem statement. We note that this is the first $\sigmatwo$-complete manipulation result.

\begin{theorem}
\label{t:sm}
Slater-Manipulation, where candidates have unary weights, is $\sigmatwo$-complete, even for one manipulator.
\end{theorem}

{\bf The Slater rule}~\cite{sla:j:slater} can be viewed as a qualitative version of Kemeny. It is defined as follows.
A ranking $>$ is a Slater consensus if the number of disagreements with the majority graph induced by the voters is minimal (note that for Slater
we look at the induced majority graph while for Kemeny we look at the
induced {\em weighted} majority graph).
In our Slater proofs, we will often look at the Slater {\em score} of a ranking, which is the number of {\em agreements} with the majority graph, i.e., $\|C\|(\|C\|-1)/2$ minus the number of disagreements. So, the higher the score, the better the ranking.

Candidates with weights for Kemeny are a natural notion~\cite{kum-vas:c:weighted-candidates}. For candidates
with weights, the contribution of each
candidate to the score is multiplied by its weight. For our result, we need only unary weights, which is a step in the direction of not needing weights.

The high-level reason that we obtain this result for Slater and not for Kemeny is that in Slater we can ``freeze'' certain arcs in the majority graph. For example, if we have three nonmanipulators all voting $a > b$, and we have one manipulator, then the manipulator cannot change the contribution to the Slater score of the pair $\{a,b\}$. Note that this is not the case for Kemeny.

Candidates with weights also give more structure to the manipulator. For example, if we have two candidates $a$ and $b$ of weight 10, then the manipulator can rank $a > b$ or $b > a$. If we replace $a$ by 10 little $a$'s and 10 little $b$'s, the manipulator can rank those in any messy order it wants.

\smallskip

\noindent
{\it Proof Sketch of Theorem~\ref{t:sm}.}
To show $\sigmatwo$-hardness, we will reduce from QSAT$_2$~\cite{sto:j:poly,wra:j:complete}. 
Consider cnf formula $\phi = D_1 \wedge \dots \wedge D_{m-1}$ over variables $x_2, \ldots, x_n$ and let
$\phi' = (x_1 \vee D_1) \wedge \dots \wedge (x_1 \vee D_{m-1}) \wedge \neg x_1$.
(Notice that $\phi'$ has $m$ clauses over variables $x_1, \dots, x_n$.) Without loss of generality, assume that if $\phi$ is not satisfiable, then at most $m-3$ clauses can be satisfied (this can be accomplished by doubling each clause).
We will in polynomial time compute an election with one manipulator such that 
$\exists x_{n'+1} \cdots x_n 
\neg (\exists x_2 \cdots x_{n'} \phi(x_2, \dots , x_n))$ if and only if the manipulator can vote such that the candidate $+_1$ becomes a winner.

First note that if $\exists x_{n'+1} \cdots x_{n} \neg (\exists x_2 \cdots x_{n'} \phi(x_2, \dots , x_n))$, then $\exists x_{n'+1} \cdots x_{n}$ such that any assignment with $x_1$ = true satisfies $m-1$ clauses of $\phi'$ and any assignment with $x_1$ = false satisfies at most $m-2$ clauses of $\phi'$. 
If it is not the case that $\exists x_{n'+1} \cdots x_{n} \neg (\exists x_2 \cdots x_{n'} \phi(x_2, \dots , x_n))$, then any assignment with $x_1$ = true satisfies $m-1$ clauses of $\phi'$ and there is an assignment with $x_1$ = false that satisfies $m$ clauses of $\phi'$.

Now apply the reduction from MAX-SAT to Slater score from~\citet{con:c:slater} to $\phi'$, with the following change. We replace each size $M$ ``super-candidate'' (a group of $M$ candidates that, for the purposes of Slater score, can be treated as one single candidate of weight $M$) by one candidate of weight $M$. This ensures that we only get Slater consensuses of a specific form and no ``rogue'' consensuses (this was not a problem in~\citet{con:c:slater}, since for the purposes of Slater scores it is enough that there exist a Slater consensus of the appropriate form; however, since we are interested in whether a specific candidate can be a winner or not, we need to preclude rogue consensuses with a rogue winner). 
This computes a tournament\footnote{For every pair of vertices
$a,b$, $a \to b$ or $b \to a$, but not both.} in which each variable $x_i$ is represented by a subtournament $T_i$
(which includes the vertices $+_i$ and $-_i$)
and each clause by a candidate $c_k$. The relevant properties of the reduction are as follows.
\begin{itemize}
 \item All Slater consensuses rank $T_1 > \cdots > T_n$.
    \item Slater consensuses correspond to assignments satisfying a maximum number of clauses of $\phi'$ in the following way. 
    For $C_k$ a true clause, 
    candidate $c_k$ is ranked (in a specific way) among the candidates in a subtournament $T_i$ whose ranking encodes an assignment to $x_i$ that makes $C_k$ true. 
    \item If $T_1$'s ranking encodes $x_1$ = true, then candidate $+_1$ is ranked first. If $T_1$'s ranking encodes $x_1$ = false, then candidate $-_1$ is ranked first.
    \item $+_1$ is a Slater winner or $-_1$ is a Slater winner.
    \item There is an assignment that satisfies $\geq k$ clauses of $\phi'$ if and only if the Slater score is $\geq B + kM$ (here, $B$ (the baseline score) and $M$ are polynomial-time computable constants that are small enough to be given in unary).
\end{itemize}
We want to keep as much of this construction as possible. First we double every voter, so that the arc weights in the induced tournament are all 2. We have one manipulator. Note that one manipulator cannot change an arc of weight 2. We will now change the tournament a little, in such a way that the manipulator can ``set'' the values of the existential variables
($x_{n'+1}, \ldots, x_{n}$), but nothing else. 

In the construction, we change how the existential variables are represented. Each such variable $x_i$ will be represented by a graph consisting of four candidates $+_i, -_i, b_i, d_i$, each of weight $M$ (recall that we allow unary weights for the candidates). These four candidates are connected by the following weight-2 arc:
\centerline{\includegraphics[scale=0.75]{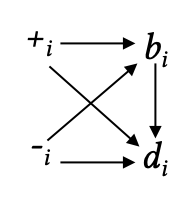}}

The only ``undeclared'' arc is between $+_i$ and $-_i$. This arc will be determined by the vote of the manipulator. $+_i > -_i$ will correspond to setting $x_i$ to true and $-_i > +_i$ will correspond to setting $x_i$ to false. Let $T'_i$ be the subtournament after the manipulator vote.

For clause candidate $c_k$, we add the following arcs.
\begin{itemize}

    \item
    If $x_i$ occurs positively in $C_k$, add arcs\\ $+_i \rightarrow c_k, c_k \rightarrow -_i, c_k \rightarrow b_i, d_i \rightarrow c_k$.
    
    \item
    If $x_i$ occurs negatively in $C_k$, add arcs\\ $-_i \rightarrow c_k, c_k \rightarrow +_i, c_k \rightarrow b_i, d_i \rightarrow c_k$.
    
    \item
    If $x_i$ does not occur in $C_k$, add arcs\\ $c_k \rightarrow +_i, c_k \rightarrow -_i, b_i \rightarrow c_k, d_i \rightarrow c_k$. 
\end{itemize}

All other arcs are unchanged.
In particular, all Slater consensuses rank $T_1 > \cdots > T_{n'} > T'_{n'+1} > \cdots > T'_n$.
Note that if we rank candidate $c_k$ before or after $T'_i$, this contributes a baseline score of $2M$ to the Slater score. The only way a clause candidate $c_k$ can gain points from $T'_i$ over the baseline score of $2M$ is if $c_k$ is ranked among the candidates in $T'_i$ and the value of $x_i$ encoded by the ranking of $T'_i$ makes $C_k$ true.
In that case, we gain $M$ extra points.

\begin{example}
For example, if $x_i$ is true and $x_i$ occurs positively in $C_k$, we obtain the subtournament below and we can order $+_i > c_k > -_i > b_i > d_i$ so that $c_k$ gains $3M$ points from $T'_i$ for the Slater score.

\centerline{\includegraphics[scale=0.75]{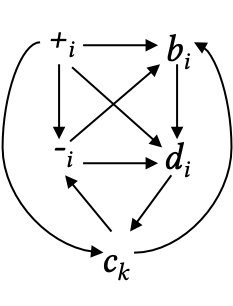}}

\end{example}

From this, we get the following, for a specific fixed assignment to $x_{n'+1}, \ldots, x_{n}$ (and the manipulator voting accordingly).

\begin{itemize}
    \item Slater consensuses correspond to assignments satisfying a maximum number of clauses of $\phi'$ in the following way. 
    For $C_k$ a true clause, 
    candidate $c_k$ is ranked (in a specific way) among the candidates in a subtournament $T_i$ or $T'_i$ whose ranking encodes an assignment to $x_i$ that makes $C_k$ true. 
    \item If $T_1$'s ranking encodes $x_1$ $=$ true, then $+_1$ is ranked first. If it encodes $x_1$ $=$ false, then $-_1$ is ranked first.
    \item $+_1$ is a Slater winner or $-_1$ is a Slater winner.
    \item There is an assignment that satisfies $\geq k$ clauses of $\phi'$ if and only if the Slater score is $\geq \widehat{B} + kM$ (here, $\widehat{B}$ is the baseline score of the new construction).
\end{itemize}

If $\exists x_{n'+1} \cdots x_{n} \neg (\exists x_2 \cdots x_{n'} \phi(x_2, \dots , x_n))$, then $\exists x_{n'+1} \cdots x_{n}$ such that any assignment with $x_1$ = true satisfies $m-1$ clauses of $\phi'$ and any assignment with $x_1$ = false satisfies at most $m-2$ clauses of $\phi'$. Let the manipulator vote according to the assignment to $x_{n'+1} \cdots x_{n}$. Then the Slater score of a total order
starting with $-_1$ is $< \widehat{B}+(m-1)M$ and the Slater score of a total order starting with $+_1$ is $\geq \widehat{B} + (m-1)M$. It follows that $+_1$ is a Slater winner.

For the converse, suppose the manipulator can vote such that $+_1$ is a winner. Consider the assignment to $x_{n'+1} \cdots x_{n}$ induced by the manipulator. If $\phi(x_2, \dots , x_{n'})$ were satisfiable, then any assignment with $x_1$ = true satisfies $m-1$ clauses of $\phi'$ and there is an assignment with $x_1$ = false that satisfies $m$ clauses of $\phi'$. It follows that the Slater score $\geq \widehat{B} + mM$ and that the ranking of $T_1$ in any Slater consensus encodes that $x_1$ is false. This implies that $-_1$ is always ranked first, which contradicts the assumption that $+_1$ is not a winner.\hspace*{1em}\literalqed\bigskip

Slater is an interesting system in itself (see, e.g.,~\citet{hul-fur:c:pairwise-preference-learning}
for motivation from the preference learning literature). But here we are mostly interested in the closeness of Slater to Kemeny, to strengthen the evidence of Theorem~\ref{t:sm} that Kemeny-Manipulation is $\sigmatwo$-complete.

Many lower bound proofs for Kemeny transfer to Slater and vice versa by the following simple observation (this is implicit in any source comparing Kemeny and Slater and explicitly stated for tournaments where every arc has weight 1 in~\citet{bac-bra-gei-har-kar-pet-see:j:takes-a-few}).

\begin{observation}
\label{o:ks}
If all weights in the weighted majority graph are the same, then the Kemeny consensus and Slater consensuses coincide.
\end{observation}

Looking back at the proofs of the results from the previous section,
we immediately obtain the following corollaries.

\begin{corollary}
\label{c:SCR}
Slater Consensus Recognition is \conp-complete.
\end{corollary}

\begin{corollary}
Slater-CDC-to-Consensus is $\sigmatwo$-complete.
\end{corollary}

The definition of Slater from this section allows an even number of 
voters. Not all Slater definitions allow ties, i.e., Slater is
sometimes defined only for the case where the
majority graph is a tournament. And also Kemeny for tournaments is an 
interesting problem.
The proofs from the previous section construct elections with an even number of voters and so do not give the analogous results about tournaments.
It is much more difficult to prove hardness for tournaments. For example, feedback arc set is one of the original 21 NP-complete problems from~\citet{kar:b:reducibilities}, but the complexity of feedback arc set for tournaments was open for a long time. The NP-hardness for feedback arc set for tournaments was shown by \citet{ail-cha-new:j:kemeny-approx}, whose
approach was derandomized by
\citet{alo:j:ranking-tournaments}. \citet{con:c:slater} gave a direct proof of the result.
We will modify the lovely reduction from~\citet{con:c:slater} to prove the following. For Slater this answers an open question from \citet{hud:j:slater}.
For details see the appendix.

\begin{theorem}
\label{t:CRT}
Slater and Kemeny Consensus Recognition for tournaments is \conp-complete.
\end{theorem}

\section{Manipulation-to-Consensus}
\label{sec:borda}

Recall from Observation~\ref{o:km} that for Kemeny-Manipulation-to-Consensus the optimal action for the 
manipulators is to vote their desired consensus.
In contrast we show that for Borda-Manipulation-to-Consensus
it is hard to compute the optimal action for the manipulators.
The Borda election system~\cite{bor:j:borda-paper} is an important rule that can be used to produce a consensus by ranking each candidate by their Borda score.
For an $m$-candidate election, each voter contributes $m-i$ points to the candidate ranked $i$th in their
vote. Note that in a Borda consensus candidates with the same score are tied.

We first show that for Borda it is not always the case that a manipulator should vote the desired consensus.

\begin{example}\label{ex:borda-consensus}
Let there be the following
five nonmanipulative voters: Two voters voting $a > b > c > d$, two voters voting $b > a > c > d$, and one voter voting $b > c > a > d$.
Let there be one manipulator with a preferred consensus of $a > b > c > d$.

Before manipulation, the candidates have the following Borda
scores: $\score{a} = 11, \score{b} = 13, \score{c} = 6,\ {\rm and}\ \score{d} = 0$, and so the consensus
is $b > a > c >~d$.

If the manipulator votes their preferred consensus
the resulting Borda scores are:
$\score{a} = 14, \score{b} = 15, \score{c} = 7,\ {\rm and}\ \score{d} = 0$, with the Borda consensus of $b > a > c > d$.

However, manipulation is possible when the manipulator instead votes $a > c > d > b$.
\end{example}

We now consider the complexity of Borda-Manipulation-to-Consensus.
The proof from~\citet{dav-kat-nar-wal-xia:j:borda-manip}, which shows that
coalitional manipulation for Borda is NP-complete constructs an election such that manipulation is possible if and
only if after manipulation the candidates $p, a_1, \ldots, a_{q+1}$ are
all tied with the highest Borda score and the remaining candidate $a_{q+2}$
has a strictly lower score, i.e., the Borda consensus
is  $\{p, a_1, \dots, a_{q+1}\} > a_{q+2}$.
It follows that:

\begin{theorem}
Borda-Manipulation-to-Consensus is \np-complete.
\end{theorem}

This immediately implies that the optimal action for the manipulators is not polynomial-time computable, unless P~=~NP.

\section{Conclusion}
We showed that even checking if a given ranking is a Kemeny consensus is coNP-complete. We also showed that, though determining whether a ranking is a Kemeny consensus is hard, the optimal action for the manipulators to reach a consensus is easy. We provided evidence that this simplicity is caused by the combination of election system (Kemeny), manipulative action (manipulation), and manipulative goal (consensus).

For future work, we are most interested in showing our conjecture that Kemeny-Manipulation(-to-Winner) is $\sigmatwo$-complete. In addition, the study of elections where candidates have weights (unary or even binary) is very natural and interesting.

\section*{Acknowledgements}
This work was supported in part by NSF-DUE-1819546. Research done in part while Zack Fitzsimmons was on
research leave at Rensselaer Polytechnic Institute.
We thank the reviewers for their helpful feedback and suggestions.

\appendix

\section{Appendix} %

\subsection*{Details for Observation~\ref{o:km}}

Observation~\ref{o:km} follows immediately from the statement below, which implies that for $X$ a consensus, if we replace a manipulator vote $\mu$ by $X$, $X$ is still a consensus.

For rankings $X$ and $\widehat{X}$, $V$ a collection of votes, and $\mu$ a vote, let $V_\mu = V \cup \{\mu\}$  and let $V_X = V \cup \{X\}$. We will show that
$$\dist{V_\mu, X} \leq \dist{V_\mu, \widehat{X}} \Rightarrow \dist{V_X,X} \leq \dist{V_X, \widehat{X}}.$$
Where $\dist$ denotes the Kendall tau distance.

We can rewrite $\dist{V_\mu, X}$ as $\dist{V_X, X} + \dist{\mu, X}$ (since $\dist{X,X} = 0$), and we can rewrite
$\dist{V_\mu, \widehat{X}}$ as $\dist{V_X, \widehat{X}} - \dist{X, \widehat{X}} + \dist{\mu, \widehat{X}}$. And so we now have
$$\dist{V_X, X} + \dist{\mu, X}
\leq \\ \dist{V_X, \widehat{X}} - \dist{X, \widehat{X}} + \dist{\mu, \widehat{X}}.$$
We know that $\dist{\mu, X} + \dist{X,\widehat{X}} \ge \dist{\mu, \widehat{X}}$ by the triangle equality.
This gives us the following.
$$\dist{V_X, \widehat{X}} - \dist{X, \widehat{X}} + \dist{\mu, \widehat{X}} - \dist{\mu,X} \leq \dist{V_X, \widehat{X}}.$$
Therefore
$\dist{V_X, X} \leq \dist{V_X, \widehat{X}}.$

\subsection*{Proof of Theorem~\ref{t:mgndr}}

We will show that Minimum GND Recognition is $\pitwo$-complete, giving
the first completeness result at the second level of the polynomial hierarchy for optimal solution recognition.

Recall that Minimum GND Recognition is defined as follows.

\prob{Minimum GND Recognition}
{A graph $G$, integer $\ell$, and set of vertices $X$.}%
{Is $X$ is minimum set of vertices such that $G - X$ does not contain $K_{\ell+1}$?}

We will show the following equivalent problem $\pitwo$-complete. 

\prob{Minimum GND$'$ Recognition}
{A graph $G$, integer $\ell$, and set of vertices $X$.}%
{Is $X$ is minimum set of vertices such that $G - X$ does not contain an independent set of size $\ell+1$?}

\begin{theorem}\label{t:mgndprimer}
Minimum GND$'$ Recognition is $\pitwo$-complete.
\end{theorem}

Note that these problems are clearly and simply equivalent, since $(G,\ell,X)$ is in Minimum GND Recognition if and only if $(\overline{G},\ell,X)$ is in Minimum GND$'$ recognition. However, since we will be modifying the NP-hardness reduction to Vertex Cover, it is much easier to think about independent sets than cliques. In fact, this change of perspective also gives a simpler reduction and proof of the $\sigmatwo$-hardness of Generalized Node Deletion~\cite{rut:j:prop-truth-maintenance}.

To show $\pitwo$-hardness, we will reduce the $\sigmatwo$-complete QSAT$_2$ problem~\cite{sto:j:poly,wra:j:complete} to the complement of Minimum GND$'$ Recognition. 
Consider the formula $\exists x_1 \cdots x_n \neg (\exists y_1 \cdots y_{n} \phi(x_1, \dots , x_n,y_1, \ldots, y_n))$, where $\phi$ is in 3cnf. Now apply to standard reduction from 3SAT to 
Vertex Cover~\cite{kar:b:reducibilities}
on $\phi$, except that we do not have an edge between $x_i$ and $\overline{x_i}$.
The gives the following graph $G$ on $4n + 3m$ vertices, where $m$ is the number of clauses of $\phi$.
\begin{itemize}
    \item For each variable $x_i$, we have two vertices $x_i$ and $\overline{x_i}$.
 \item For each variable $y_i$, we have two vertices $y_i$ and $\overline{y_i}$ connected by an edge.
  \item For the $i$th clause of $\phi$, we have a triangle consisting of three vertices $a_i$, $b_i$, and $c_i$.
    \item If the $i$th clause of $\phi$ is $\ell_1 \wedge \ell_2 \wedge \ell_3$, then connect $a_i$ to $\ell_1$, $b_i$ to $\ell_2$ and $c_i$ to $\ell_3$.
\end{itemize}
Let $\alpha_1 \cdots \alpha_n$ be an assignment to $x_1 \cdots x_n$. Let 
$G_\alpha$ be the graph obtained from $G$ by deleting the vertices corresponding to the assignment, i.e.,
$G_\alpha = G - \{x_i \ | \ \alpha_i = 1\} - \{\overline{x_i} \ | \ \alpha_i = 0\}$. Note that $G_\alpha$ has $3n + 3m$ vertices.

From the proof of~\citet{kar:b:reducibilities}, it is immediate that $\phi(\alpha_1, \ldots, \alpha_n, y_1, \ldots, y_n)$ is satisfiable if and only if $G_\alpha$ has a vertex cover of size $n+2m$ (i.e., if and only if $G_\alpha$ has an independent set of size $2n + m$). In addition, $G_\alpha$ does not have and independent set of size $2n +m + 1$.

The ``$2n+m$'' will be the ``$\ell +1$'' in our instance of Minimum GND$'$ Recognition, i.e., $\ell = 2n + m - 1$. Next, we make sure that it is attractive to delete one of $\{x_i, \overline{x_i}\}$ for all $i$. We will modify our graph $G$ using the notion of ``forcing''
from~\citet{rut:j:prop-truth-maintenance} (but then for independent sets).
For all $i, 1 \leq i \leq n$, add $2n+m-2$ vertices $I_i$ and $2n+m-2$ vertices $I'_i$.
The vertices in $I_i$ are connected to all vertices not in $I_i \cup \{x_i, \overline{x_i}\}$ and 
the vertices in $I'_i$ are connected to all vertices not in $I'_i \cup \{x_i, \overline{x_i}\}$.
Note that $\{x_i, \overline{x_i}\} \cup I_i \cup I'_i$ consists of two independent sets of size $2n+m$ with intersection $\{x_i,\overline{x_i}\}$. We can decrease the size of both independent sets by removing one vertex if only if that vertex is $x_i$ or $\overline{x_i}$.
Call the thus-padded graph $H$ and let $H_\alpha$ be $H$ with the $n$ vertices corresponding to assignment $\alpha$ deleted. The size $2n + m$ independent sets of $H$ are:
\begin{itemize}
\item $I_i \cup \{x_i,\overline{x_i}\}$,
\item $I'_i \cup \{x_i,\overline{x_i}\}$, and
\item all independent sets of size $2n + m$ of $G$.  
\end{itemize}
The independent sets of size $2n+m$ of $H_\alpha$ are exactly the independent sets of 
size $2n+m$ of $G_\alpha$, and so 
$\phi(\alpha_1, \ldots, \alpha_n, y_1, \ldots, y_n)$ is not satisfiable if and only if
$H_\alpha$ does not contain an independent set of size $2n + m$. In addition, 
$H_\alpha$ does not contain an independent set of size $2n + m + 1$.

Now we show that\\ 
$\exists x_1 \cdots \exists x_n \neg (\exists y_1 \cdots y_{n} \phi(x_1, \dots , x_n,y_1, \ldots, y_n))$  if and only if $X = \{x_1, \ldots, x_n\} \cup \{\overline{x_1}\}$ is not a minimum set of vertices such that $H - X$ does not contain an independent set of size $2n+m$, which completes the proof.

\begin{description}
\item[($\Rightarrow$)] Let $\alpha_1 \cdots \alpha_n$ be such that
$\phi(\alpha_1, \ldots, \alpha_n, y_1, \ldots, y_n)$ is not satisfiable.
Then $H_\alpha$ does not contain an independent set of size $2n+m$.
It follows that the set of $n$ vertices corresponding to the assignment is a solution of size $n$ and thus
$X$ is not an optimal solution.

\item[($\Leftarrow$)] Since $H - \{x_1, \ldots, x_n\}$ does not contain an independent set of size $2n+m+1$ and any independent set of $H - \{x_1, \ldots, x_n\}$ of size $2n+m$ contains $\overline{x_1}$ (this is because any vertex cover of $H - \{x_1, \ldots, x_n\}$ of size $n + 2m$ must consist of exactly one of $\{y_i,\overline{y_i}\}$ and exactly two vertices of each clause triangle), it follows that $H - X$ does not contain an independent set of size $2n+m$. Thus, $X$ is a solution, but not an optimal solution.
It follows that there is a set $X'$ of $n$ vertices such that $H - X'$ does not have an independent set
of size $2n+m$. The only way to decrease the size of the $2n$ independent sets of size 
$2n+m$ of the form $I_i \cup \{x_i,\overline{x_i}\}$ and $I'_i \cup \{x_i,\overline{x_i}\}$ is to delete exactly
one of $x_i$ and $\overline{x_i}$. Let $\alpha$ be the corresponding assignment. Then
$H - X' = H_\alpha$ and $H_\alpha$ does not have an independent set of size
$2n + m$, which implies that
$\phi(\alpha_1, \ldots, \alpha_n, y_1, \ldots, y_n)$ is not satisfiable.
\end{description}

\subsection*{Minimum Vertex Cover and FAS Recognition Deletion}

We show that the natural deleting analogues of
Minimum Vertex Cover and FAS Recognition are $\sigmatwo$-complete.

\prob{Minimum Vertex Cover Recognition Deletion}%
{A graph $G= (V,E)$, delete limit $k$, and set of vertices~$X$.}%
{Does there exist a set $W \subseteq V$ of at most $k$ vertices such that $X$ is a minimum vertex cover of $G-W$?\footnote{Note that this implies that $W \subseteq V - X$.}}

\begin{theorem}
\label{t:mvcrd}
Minimum Vertex Cover Recognition Deletion is $\sigmatwo$-complete.
\end{theorem}

\begin{proofs}
As is typical, the $\sigmatwo$ upper bound is immediate. $\sigmatwo$ lower bounds are often hard to prove, in part because there are fewer known $\sigmatwo$-complete problems (see~\citet{sch-uma:j:PH-part-one} for a list)
and also because one needs a closer correspondence between the two problems than for NP-hardness reductions. A $\sigmatwo$-complete problem closely related to ours is Vertex-Cover-Member-Select~\cite{fit-hem-hoo-nar:c:vhc}.

\prob{Vertex-Cover-Member-Select}%
{A graph $G= (V,E)$, set $V' \subseteq V$ of deletable vertices, delete limit $k$, and vertex $\hat{v} \in V$.}%
{Does there exist a set $W \subseteq V'$ of at most $k$ vertices such that $\hat{v}$ is a member of a minimum vertex cover of $G-W$?}

This problem does not easily reduce to ours. But by amazing luck, the reduction establishing $\sigmatwo$-hardness for Vertex-Cover-Member-Select ``builds in'' a specific fixed vertex cover that contains $\hat{v}$. This is exactly what we need to establish the $\sigmatwo$-hardness of Minimum Vertex Cover Recognition Deletion. To be explicit, in the construction of the proof of~\cite[Lemma 4]{fit-hem-hoo-nar:c:vhc}, take $k=n$ and $X = \{\hat{v}\} \cup \{y_1, y_2, \ldots , y_n\} \cup \{c_{i,j} \ | \ 1 \leq i \leq m, 1 \leq j \leq 3\})$.
\end{proofs}

The generalization of the standard reduction from Karp~\shortcite{kar:b:reducibilities} from the proof of Theorem~\ref{t:minfas} generalizes in the obvious way to a reduction from Minimum Vertex Cover Recognition Deletion to the following problem.

\prob{Minimum FAS Recognition Deletion}%
{An irreflexive and antisymmetric directed graph $G = (V,A)$, delete limit $k$, and a set of arcs $X$.}%
{Does there exist a set $W \subseteq V$ of at most $k$ vertices such that $X$ is a minimum fas of $G-W$?}

\begin{theorem}\label{t:mfasrd}
Minimum FAS Recognition Deletion
is $\sigmatwo$-complete.
\end{theorem}

\begin{proofs}
We reduce Minimum Vertex Cover Recognition Deletion to Minimum FAS Recognition Deletion
as follows. For any graph $H$, define directed graph $\widehat{H}$ as in the construction of the reduction from Vertex Cover to FAS~\cite{kar:b:reducibilities}, i.e., 
\begin{itemize} 
\item
$V(\widehat{H}) = \{v,v' \ | \ v \in V(H)\}$, and
\item $A(\widehat{H}) = \{(v,v') \ | \ v \in V(H)\} \cup \{(v',w),(w',v) \ | \  \{v,w\} \in E(H)\}$.
\end{itemize}
And for any set of vertices $Y$ of $H$, let $\widehat{Y} = \{(v,v') \ | \ v \in Y\}$.

Let $G$ be a graph, $k$ a delete limit, and $X$ a set of vertices of $G$. We will show that the $(G,k,X)$ is in Minimum Vertex Cover Recognition Deletion if and only if $(\widehat{G},k,\widehat{X})$ is in Minimum FAS Recognition Deletion.

It follows from the proof of the reduction from Karp~\cite{kar:b:reducibilities} that for all graphs $H$ and for all sets of vertices $Y$ of $H$, $Y$ is a minimum vertex cover of $H$ if and only if $\widehat{Y}$ is a minimum fas of $\widehat{H}$.

If $X$ is a minimum vertex cover of $G - W$, then  $\widehat{X}$ is a minimum fas of $\widehat{G - W} = \widehat{G} - \{v,v' \ | \ v \in W\}$. Since the only arc into $v'$ is $(v,v')$, it follows that $\widehat{X}$ is a minimum fas of $\widehat{G} - W$.  

For the converse, suppose that $\widehat{X}$ is a minimum fas of $\widehat{G} - W'$. Let $W = \{v \ | \ v \in W \mbox{ or } v' \in W\}$ (note that $\|W\| \leq \|W'\|)$.
Then $\widehat{X}$ is a minimum fas of $\widehat{G - W}$, and it follows that $X$ is a minimum vertex cover of $G - W$. 
\end{proofs}

\subsection*{Proof of Theorem~\ref{t:kcdcc}}

As mentioned in the core text, 
to prove $\sigmatwo$-completeness of Kemeny-CDC-to-Consensus, we need $\sigmatwo$-complete versions of Vertex Cover and Feedback Arc Set that look more like Kemeny-CDC-to-Consensus. In particular, we need to make sure that the solutions for Vertex Cover and Feedback Arc Set are (not necessarily optimal) solutions for the whole graph. 

\prob{Minimum Vertex Cover Recognition Restriction}%
{A graph $G= (V,E)$, delete limit $k$, and minimal vertex cover $X$ of $G$.}%
{Does there exist a set $W \subseteq V$ of at most $k$ vertices such that $X - W$ is a minimum vertex cover of $G-W$?}

\begin{theorem}
\label{t:mvcrr}
Minimum Vertex Cover Recognition Restriction is $\sigmatwo$-complete.
\end{theorem}

\begin{proofs}
We reduce from the following \sigmatwo-complete problem~\cite{rut:j:prop-truth-maintenance}.

\prob{Generalized Node Deletion}%
{A graph $G= (V,E)$ and integers $k$ and $\ell$.}%
{Does there exist a set $W \subseteq V$ of at most $k$ vertices such that $G-W$ does not contain $K_{\ell+1}$?}

For the reduction, map $(G,k,\ell)$ to graph $H = \overline{G} + \overline{K_\ell}$ (where $+$ denotes
the join), delete limit $k$, and $X = V(G)$. 
It is immediate that $X$ is a minimal vertex cover of $H$.
To show that the reduction is correct, first assume that there is a set $W \subseteq V(G)$ of at most $k$ vertices such that $G-W$ does not contain $K_{\ell+1}$ (i.e., the size of a maximum independent set of $\overline{G}-W$ is at most $\ell$). Now consider $H - W$. Note that there are two types of  minimal vertex covers of $H-W$. 
\begin{enumerate}
  \item $V(G) - W$ (of size $\|V(G)\| - \|W\|$).
  \item $V(\overline{K_{\ell}})$ unioned with a minimum vertex cover of $\overline{G} - W$. Since a minimum vertex cover is the complement of a maximum independent set, and the size of a maximum independent set of $\overline{G} - W$ is at most $\ell$, it follows that the size of minimum vertex covers of this kind is at least $\ell + \|V(G)\| - \|W\| - \ell = \|V(G)\| - \|W\|$.
\end{enumerate}
It follows that $V(G) - W = X - W$ is a minimum vertex cover of $H - W$.

For the converse, suppose that we can delete a set $W$ of at most $k$ vertices from $H$ such that $X - W = V(G) - W$ is a minimum vertex cover of $H - W$. Suppose for a contradiction that $G - W$ contains $K_{\ell+1}$. Then $\overline{G} - W$ contains $\overline{K_{\ell+1}}$ and so $\overline{G} - W$ contains a minimum vertex cover of size $\leq \|V(G) - W\| - (\ell+1)$. Combined with $V(\overline{K_\ell}) - W$, this gives a minimum vertex cover of $H - W$ of size
$< \|V(G) - W\|$. But that contradicts the fact that $V(G) - W$ is a minimum vertex cover of $H - W$.
\end{proofs}

The generalization of the reduction from Karp~\shortcite{kar:b:reducibilities} used in the proof of Theorem~\ref{t:minfas} generalizes in the obvious way to a reduction from Minimum Vertex Cover Recognition Restriction to the following problem. 

\prob{Minimum FAS Recognition Restriction}%
{An irreflexive and antisymmetric directed graph $G = (V,A)$, delete limit $k$, and a minimal fas $X$ of $G$.}%
{Does there exist a set $W \subseteq V$ of at most $k$ vertices such that $X \cap ((V - W) \times (V - W))$
is a minimum fas of $G-W$?
}

\begin{theorem}
\label{t:mfrr}
Minimum FAS Recognition Restriction is $\sigmatwo$-complete.
\end{theorem}

\begin{proofs}
We reduce Minimum Vertex Cover Recognition Restriction to Minimum FAS Recognition Restriction
as follows. 

As in the proof of Theorem~\ref{t:mfasrd}, we will use the construction from~\cite{kar:b:reducibilities}, using the notation from the proof of Theorem~\ref{t:mfasrd}.
Let $G$ be a graph, $k$ a delete limit, and $X$ a minimal vertex cover of $G$. 
We will
show that $(G,k,X)$ is in Minimum Vertex Cover Recognition Restriction if and only if
$(\widehat{G},k,\widehat{X})$ is in Minimum FAS Recognition Restriction.
 
First note that $\widehat{X}$ is a minimal fas of $\widehat{G}$.

Suppose that $X - W$ is a minimum vertex cover of $G - W$. Then $\widehat{X - W}$ is a minimum fas of $\widehat{G - W} = \widehat{G} - \{v,v' \ | \ x \in W\}$. Since the only arc into $v'$ is $(v,v')$, it follows that $\widehat{X - W}$ is a minimum fas of $\widehat{G} - W$. In addition, $\widehat{X - W} = \widehat{X} \cap ((V(\widehat{G}) - W) \times (V(\widehat{G}) - W))$.

For the converse, suppose that $\widehat{X} \cap ((V(\widehat{G}) - W') \times (V(\widehat{G}) - W'))$ is a minimum fas of $\widehat{G} - W'$. Let $W = \{v \ | \ v \in W \mbox{ or } v' \in W\}$ (note that $\|W\| \leq \|W'\|)$. Then $\widehat{X - W} = \widehat{X} \cap ((V(\widehat{G}) - W') \times (V(\widehat{G}) - W')$ and $\widehat{X - W}$ is a minimum fas of $\widehat{G - W}$. It follows that $X - W$ is a minimum vertex cover of $G - W$.
\end{proofs}

To finish the proof of Theorem~\ref{t:kcdcc}, we reduce
Minimum FAS Recognition Restriction to Kemeny-CDC-to-Consensus,
using the reduction from the proof of Theorem~\ref{t:KCR}, keeping the delete limit the same. 
Given $(G = (C,A), k, X)$, if $X$ is not a minimal fas
then output a string that is not in Kemeny-CDC-to-Consensus. If $X$ is a minimal fas, then output $e(G)$, where $e(G)$ is the election corresponding to $G$ from Lemma~\ref{l:faskc}, delete limit $k$, and a total order $\widehat{X}$ consistent with $G - X$.

Now suppose there exists a set $D$ of at most $k$ vertices such that $X' = X \cap ((C-D) \times (C-D))$ is a minimum fas of $G - D$.
Then $\widehat{X}$ restricted to $C - D$ is consistent
with $G - X'$ and it follows from Lemma~\ref{l:faskc}
that $\widehat{X}$ restricted to $C - D$ is a Kemeny consensus of $e(G-D)$.
Since $e(G-D)$ is $e(G)$ restricted to $C - D$,
it follows that $(e(G),k,\widehat{X})$ is in Kemeny-CDC-to-Consensus.

For the converse, let $D$ be a set of at most $k$ candidates such that $\widehat{X}$ restricted to $C - D$ is a Kemeny consensus of $e(G)$ restricted to $C - D$.
Let $X' = X \cap ((C - D) \times (C - D))$.
Then $X'$ is a minimal fas of $G - D$ (since $X$ is a minimal fas of $G$) and $\widehat{X}$ restricted to $C - D$ is consistent with $G - X'$ and $e(G)$ restricted to $C - D$ is $e(C-D)$. It follows from Lemma~\ref{l:faskc} that $X'$ is a minimum fas of $G - D$, and so $(G,k,X)$ is in 
Minimum FAS Recognition Restriction.

\subsection*{Details for Theorem~\ref{t:CRT}}
\begin{proofsketch}
The coNP upper bound is immediate.
To show hardness, we modify and use the reduction (from MAX-SAT) due to Conitzer~\shortcite{con:c:slater}.
Given a cnf formula $C_1 \wedge \dots \wedge C_m$ over variables $x_1, \ldots, x_n$, this reduction computes in polynomial time a tournament $T$ with the following properties, where $B$ (the baseline score) and $M$ are polynomial-time computable constants and $M > m(m-1)/2$.
\begin{enumerate}
    \item If there exists an assignment that satisfies $k$ clauses, then there exists a total order with Slater score $\geq B + kM$.
    \item If there is a total order with Slater score $>B+ (k-1)M + m(m-1)/2$, then there is an assignment that satisfies at least $k$ clauses.
\end{enumerate}

To show that Slater Consensus Recognition for tournaments is coNP-hard, reduce from the
complement of SAT.
Consider cnf formula $\phi = D_1 \wedge \dots \wedge D_{m-1}$ over variables $x_2, \ldots, x_n$ and let
$\phi' = (x_1 \vee D_1) \wedge \dots \wedge (x_1 \vee D_{m-1}) \wedge \neg x_1$.
(Notice that $\phi'$ has $m$ clauses over variables $x_1, \dots, x_n$.)  Apply the reduction from~\cite{con:c:slater} to $\phi'$. Note that if $\phi$ is satisfiable, then so is $\phi'$, in which case there exists a total order with Slater score $\geq B + mM$. If $\phi$ is not satisfiable, then neither is $\phi'$, but we can satisfy the first $m-1$ clauses of $\phi'$ by setting $x_1$ to true. From the properties of the reduction from~\cite{con:c:slater}, we know that the score of a Slater consensus is between $B + (m-1)M$ and $B + (m-1)M + m(m-1)/2$. But we need an explicit Slater consensus. The reason for the possible range of scores in the reduction from~\cite{con:c:slater} is that there are $m(m-1)/2$ tournament arcs that do not have to be specified to make that reduction (that shows that deciding whether the Slater score is at least $\ell$) work, namely the $m(m-1)/2$ arcs between candidates $c_1, \ldots, c_m$ that correspond to the $m$ clauses. 
We will fix these arcs by having an arc from $c_i$ to $c_j$ if and only if $i < j$. By inspection of the reduction from~\cite{con:c:slater},\footnote{
We claim that $\phi$ is not satisfiable if and only if the order $X$
    that corresponds to the assignment that sets all variables to true and starts like this: $+_1 > c_1 > \dots > c_m$ is a Slater consensus.

If $\phi$ is satisfiable, then $\phi'$ is satisfiable. Then there exists a total order with Slater score $\geq B + mM$.
But $\phi'$ is never satisfiable when $x_1$ is true, and so any total order that has $+_1$ first has Slater score at most $B + (m-1)M + m(m-1)/2 < B + mM$.
It follows that $X$ is not a Slater consensus.

If $\phi$ is not satisfiable, then $\phi'$ is not satisfiable and so the Slater score of the tournament is at most $B + (m-1)M + m(m-1)/2$. We will show that order $X$ has exactly that Slater score and so $X$ is a Slater consensus. Note that when all variables are set to true, we satisfy $m-1$ of the $m$ clauses of $\phi'$. Also note that all of these $m-1$ clauses are satisfied by $x_1$. This sets the Slater score of $X$ to $B + (m-1)M$ plus the number of arcs between $c_i$'s that are consistent with the tournament. But note that the order of the $c_i$'s in  $X$ is completely consistent with the tournament, and this adds $m(m-1)/2$ points to the Slater score.} a ranking that corresponds to an assignment for $\phi'$ that sets all variables to true and that places each $c_i, i < m$ with (the part of the ranking corresponding to) a variable that makes the corresponding clause true is a ranking with score $\geq B + (m-1)M$. Since setting $x_1$ to true makes the first $m-1$ clauses true, we can place $c_1 > c_2 > \cdots > c_m$ with (the part of the ranking corresponding to) variable $x_1$. This adds $m(m-1)/2$ points to the Slater score and so this ranking is a Slater consensus.

The same construction works for Kemeny, using Observation~\ref{o:ks}.~\end{proofsketch}


\begin{thebibliography}{36}
\providecommand{\natexlab}[1]{#1}
\providecommand{\url}[1]{\texttt{#1}}
\expandafter\ifx\csname urlstyle\endcsname\relax
  \providecommand{\doi}[1]{doi: #1}\else
  \providecommand{\doi}{doi: \begingroup \urlstyle{rm}\Url}\fi

\bibitem[Ailon et~al.(2008)Ailon, Charikar, and
  Newman]{ail-cha-new:j:kemeny-approx}
N.~Ailon, M.~Charikar, and A.~Newman.
\newblock Aggregating inconsistent information: {R}anking and clustering.
\newblock \emph{JACM}, 55\penalty0 (5):\penalty0 Article~23, 2008.

\bibitem[Alon(2006)]{alo:j:ranking-tournaments}
N.~Alon.
\newblock Ranking tournaments.
\newblock \emph{SIDMA}, 20\penalty0 (1--2):\penalty0 137--142, 2006.

\bibitem[Armstrong and Jacobson(2003)]{arm-jac:j:global-verification}
D.~Armstrong and S.~Jacobson.
\newblock Studying the complexity of global verification for {NP}-hard discrete
  optimization problems.
\newblock \emph{J.~Glob.~Optim.}, 27:\penalty0 83--96, 2003.

\bibitem[Babai(2016)]{Babai}
L.~Babai.
\newblock Graph isomorphism in quasipolynomial time [extended abstract].
\newblock In \emph{Proc.~of STOC-16}, pages 684--697. {ACM}, June 2016.

\bibitem[Bachmeier et~al.(2019)Bachmeier, Brandt, Geist, Harrenstein, Kardel,
  Peters, and Seedig]{bac-bra-gei-har-kar-pet-see:j:takes-a-few}
G.~Bachmeier, F.~Brandt, C.~Geist, P.~Harrenstein, K.~Kardel, D.~Peters, and
  H.~Seedig.
\newblock \emph{k}-{M}ajority digraphs and the hardness of voting with a
  constant number of voters.
\newblock \emph{JCSS}, 105:\penalty0 130--157, 2019.

\bibitem[{{Bartholdi}} et~al.(1989{\natexlab{a}}){{Bartholdi}}, Tovey, and
  Trick]{bar-tov-tri:j:manipulating}
J.~{{Bartholdi}}, III, C.~Tovey, and M.~Trick.
\newblock The computational difficulty of manipulating an election.
\newblock \emph{SCW}, 6\penalty0 (3):\penalty0 227--241, 1989{\natexlab{a}}.

\bibitem[{{Bartholdi}} et~al.(1989{\natexlab{b}}){{Bartholdi}}, Tovey, and
  Trick]{bar-tov-tri:j:who-won}
J.~{{Bartholdi}}, III, C.~Tovey, and M.~Trick.
\newblock Voting schemes for which it can be difficult to tell who won the
  election.
\newblock \emph{SCW}, 6\penalty0 (2):\penalty0 157--165, 1989{\natexlab{b}}.

\bibitem[{{Bartholdi}} et~al.(1992){{Bartholdi}}, Tovey, and
  Trick]{bar-tov-tri:j:control}
J.~{{Bartholdi}}, III, C.~Tovey, and M.~Trick.
\newblock How hard is it to control an election?
\newblock \emph{Math.~Comput.~Model}, 16\penalty0 (8/9):\penalty0 27--40, 1992.

\bibitem[Betzler et~al.(2011)Betzler, Niedermeier, and
  Woeginger]{bet-nei-woe:c:board-manip}
N.~Betzler, R.~Niedermeier, and G.~Woeginger.
\newblock Unweighted coalitional manipulation under the {B}orda rule is
  {NP}-hard.
\newblock In \emph{Proc.~of IJCAI-11}, pages 55--60, August 2011.

\bibitem[Conitzer(2006)]{con:c:slater}
V.~Conitzer.
\newblock Computing {S}later rankings using similarities among candidates.
\newblock In \emph{Proc.~of AAAI-06}, pages 613--619, July 2006.

\bibitem[Davies et~al.(2014)Davies, Katsirelos, Narodytska, Walsh, and
  Xia]{dav-kat-nar-wal-xia:j:borda-manip}
J.~Davies, G.~Katsirelos, N.~Narodytska, T.~Walsh, and L.~Xia.
\newblock Complexity of and algorithms for the manipulation of {B}orda,
  {N}anson's and {B}aldwin's voting rules.
\newblock \emph{AIJ}, 217:\penalty0 20--42, 2014.

\bibitem[de~Borda(1781)]{bor:j:borda-paper}
J.-C. de~Borda.
\newblock M{\'e}moire sur les {\'e}lections au scrutin.
\newblock \emph{Histoire de l'Acad{\'e}mie Royale des Sciences}, pages
  657--664, 1781.

\bibitem[Dwork et~al.(2001)Dwork, Kumar, Naor, and
  Sivakumar]{dwo-kum-nao-siv:c:rank-aggregation}
C.~Dwork, R.~Kumar, M.~Naor, and D.~Sivakumar.
\newblock Rank aggregation methods for the web.
\newblock In \emph{Proc.~of WWW-01}, pages 613--622, March 2001.

\bibitem[Faliszewski and
  Rothe(2016)]{fal-rot:b:handbook-comsoc-control-and-bribery}
P.~Faliszewski and J.~Rothe.
\newblock Control and bribery in voting.
\newblock In F.~Brandt, V.~Conitzer, U.~Endriss, J.~Lang, and A.~Procaccia,
  editors, \emph{Handbook of Computational Social Choice}, pages 146--168.
  Cambridge University Press, 2016.

\bibitem[Faliszewski et~al.(2008)Faliszewski, Hemaspaandra, and
  Schnoor]{fal-hem-sch:c:copeland-ties-matter}
P.~Faliszewski, E.~Hemaspaandra, and H.~Schnoor.
\newblock Copeland voting: Ties matter.
\newblock In \emph{Proc.~of AAMAS-08}, pages 983--990, May 2008.

\bibitem[Faliszewski et~al.(2010)Faliszewski, Hemaspaandra, and
  Schnoor]{fal-hem-sch:c:copeland01}
P.~Faliszewski, E.~Hemaspaandra, and H.~Schnoor.
\newblock Manipulation of {Copeland} elections.
\newblock In \emph{Proc.~of AAMAS-10}, pages 367--374, May 2010.

\bibitem[Fitzsimmons et~al.(2019)Fitzsimmons, Hemaspaandra, Hoover, and
  Narv{\'a}ez]{fit-hem-hoo-nar:c:vhc}
Z.~Fitzsimmons, E.~Hemaspaandra, A.~Hoover, and D.~Narv{\'a}ez.
\newblock Very hard electoral control problems.
\newblock In \emph{Proc.~of AAAI-19}, pages 1933--1940, January/February 2019.

\bibitem[Goldreich et~al.(1991)Goldreich, Micali, and
  Wigderson]{gol-mic-wig:j:zero-knowledge}
O.~Goldreich, S.~Micali, and A.~Wigderson.
\newblock Proofs that yield nothing but their validity for all languages in
  {NP} have zero-knowledge proof systems.
\newblock \emph{JACM}, 38\penalty0 (3):\penalty0 691--729, 1991.

\bibitem[Hemaspaandra et~al.(2005)Hemaspaandra, Spakowski, and
  Vogel]{hem-spa-vog:j:kemeny}
E.~Hemaspaandra, H.~Spakowski, and J.~Vogel.
\newblock The complexity of {Kemeny} elections.
\newblock \emph{TCS}, 349\penalty0 (3):\penalty0 382--391, 2005.

\bibitem[Hudry(2010)]{hud:j:slater}
O.~Hudry.
\newblock On the complexity of {S}later's problems.
\newblock \emph{EJOR}, 203\penalty0 (1):\penalty0 216--221, 2010.

\bibitem[Hudry(2013)]{hud:j:kemeny-complexity}
O.~Hudry.
\newblock Complexity of computing median linear orders and variants.
\newblock \emph{ENDM}, 42:\penalty0 57--64, 2013.

\bibitem[H{\"u}llermeier and
  F{\"u}rnkranz(2004)]{hul-fur:c:pairwise-preference-learning}
E.~H{\"u}llermeier and J.~F{\"u}rnkranz.
\newblock Comparison of ranking procedures in pairwise preference learning.
\newblock In \emph{Proc.~of IPMU-04}, 2004.

\bibitem[Jackson et~al.(2008)Jackson, Schnable, and
  Aluru]{jac-sch-alu:j:kemeny-bioinformatics}
B.~Jackson, S.~Schnable, and S.~Aluru.
\newblock Consensus genetic maps as media orders from inconsistent sources.
\newblock \emph{TCBB}, 5\penalty0 (2):\penalty0 161--171, 2008.

\bibitem[Karp(1972)]{kar:b:reducibilities}
R.~Karp.
\newblock Reducibility among combinatorial problems.
\newblock In \emph{Proc.~of Symposium on Complexity of Computer Computations},
  pages 85--103, 1972.

\bibitem[Kemeny(1959)]{kem:j:no-numbers}
J.~Kemeny.
\newblock Mathematics without numbers.
\newblock \emph{Daedalus}, 88:\penalty0 577--591, 1959.

\bibitem[Kumar and Vassilvitskii(2010)]{kum-vas:c:weighted-candidates}
R.~Kumar and S.~Vassilvitskii.
\newblock Generalized distances between rankings.
\newblock In \emph{Proc.~of WWW-10}, pages 571--580. {ACM}, April 2010.

\bibitem[Ladner(1975)]{lad:j:np-incomplete}
R.~Ladner.
\newblock On the structure of polynomial time reducibility.
\newblock \emph{JACM}, 22\penalty0 (1):\penalty0 155--171, 1975.

\bibitem[McGarvey(1953)]{mcg:j:election-graph}
D.~McGarvey.
\newblock A theorem on the construction of voting paradoxes.
\newblock \emph{Econometrica}, 21\penalty0 (4):\penalty0 608--610, 1953.

\bibitem[Meyer and Stockmeyer(1972)]{mey-sto:c:reg-exp-needs-exp-space}
A.~Meyer and L.~Stockmeyer.
\newblock The equivalence problem for regular expressions with squaring
  requires exponential space.
\newblock In \emph{Proc.~of FOCS-72}, pages 125--129, October 1972.

\bibitem[Papadimitriou and Steiglitz(1978)]{pap-ste:j:tsp}
C.~Papadimitriou and K.~Steiglitz.
\newblock Some examples of difficult traveling salesman problems.
\newblock \emph{Oper.~Res.}, 26\penalty0 (3):\penalty0 434--443, 1978.

\bibitem[Russell(2007)]{rus:t:borda}
N.~Russell.
\newblock Complexity of control of {B}orda count elections.
\newblock Master's thesis, Rochester Institute of Technology, 2007.

\bibitem[Rutenburg(1994)]{rut:j:prop-truth-maintenance}
V.~Rutenburg.
\newblock Propositional truth maintenance systems: {C}lassification and
  complexity analysis.
\newblock \emph{AMAI}, 10\penalty0 (3):\penalty0 207--231, 1994.

\bibitem[Schaefer and Umans(2002)]{sch-uma:j:PH-part-one}
M.~Schaefer and C.~Umans.
\newblock Completeness in the polynomial-time hierarchy: Part {I}: {A}
  compendium.
\newblock \emph{SIGACT News}, 33\penalty0 (3):\penalty0 32--49, 2002.

\bibitem[Slater(1961)]{sla:j:slater}
P.~Slater.
\newblock Inconsistencies in a schedule of paired comparisons.
\newblock \emph{Biometrika}, 48\penalty0 (3/4):\penalty0 303--312, 1961.

\bibitem[Stockmeyer(1976)]{sto:j:poly}
L.~Stockmeyer.
\newblock The polynomial-time hierarchy.
\newblock \emph{TCS}, 3\penalty0 (1):\penalty0 1--22, 1976.

\bibitem[Wrathall(1976)]{wra:j:complete}
C.~Wrathall.
\newblock Complete sets and the polynomial-time hierarchy.
\newblock \emph{TCS}, 3\penalty0 (1):\penalty0 23--33, 1976.

\end{thebibliography}
\end{document}